# Exploring material compositions for synthesis using oxidation states

Maung Thway[1], Andy Paul Chen[1], Haiwen Dai[1], Jose Recatala-Gomez[1], Siyu Isaac Parker Tian[2], Ruiming Zhu[1], Wenhao Zhai[3], Fengxia Wei[3], D. V. Maheshwar Repaka[3], Tonio Buonassisi[2,5], Pieremanuele Canepa[4], Kedar Hippalgaonkar[1,3]

[1] School of Materials Science and Engineering, Nanyang Technological University, 50 Nanyang Avenue, Singapore 639798, Singapore.

[2] Low Energy Electronic Systems (LEES), Singapore-MIT Alliance for Research and Technology (SMART), 1 Create Way, Singapore 138602, Singapore.

[3] Institute of Materials Research & Engineering, Agency for Science, Technology and Research (A*STAR), Singapore 138632, Singapore.

[4] Department of Materials Science and Engineering, National University of Singapore, 9 Engineering Drive 1, Singapore 117575, Singapore.

[5] Department of Mechanical Engineering, Massachusetts Institute of Technology (MIT), 77 Massachusetts Ave., Cambridge, MA 02139, USA.

## Abstract

Recent advances in machine learning techniques have made it possible to use high-throughput screening to identify novel materials with specific properties. However, the large number of potential candidates produced by these techniques can make it difficult to select the most promising ones. In this study, we develop the oxidation state probability (OSP) method which evaluates ternary compounds based on the probability (the OSP metric) of each element to adopt the required oxidation states for fulfilling charge neutrality. We compare this model with *Roost* and the Fourier-transformed crystal properties (FTCP)-based synthesizability score. Among the top 1000 systems with the most database entries in Materials Project (MP), more than 500 systems exhibit an attested compound among the top 3 compositions when ranked by the OSP metric. We find that the OSP method shows promising results for certain classes of ternary systems, especially those containing nonmetals, s-block, or transition metals. When applied to the Cu-In-Te ternary system, an interesting system for thermoelectric applications, the OSP method predicted the synthesizability of $CuIn_3Te_5$ without prior knowledge, and we have successfully synthesized $CuIn_3Te_5$ in experiment. Our method has the potential to accelerate the discovery of novel compounds by providing a guide for experimentalists to easily select the most synthesizable candidates from an arbitrarily large set of possible chemical compositions.

# Introduction

The exploration and adoption of novel functional materials can help to alleviate numerous pressing issues in society, such as climate change and accurate viral detection in the event of a global pandemic. Traditionally, the drivers of this process have been domain expertise and trial-and-error approaches[1]. Large scale computing, publicly available materials databases and machine learning (ML) methods have accelerated the discovery of novel materials[2]. These enable the screening of vast databases[3], through which potential materials with the best materials properties or the easiest synthesis routes can be identified rapidly[4]. However, a major challenge in this approach is still the vast number of potential materials that meet the desired criteria. Recently, some researchers have developed functional descriptors relating to material properties, but rarely on the ability of synthesizing the material[5], [6], [7], [8], [9], [10]. The development of synthetic databases requires an improved accuracy in the generative algorithm. As it stands generative and inverse design algorithms tend to produce a large volume of spurious and unphysical results. The number of possible ternary compounds in the search space is staggering, with over 32 million unique combinations of three elements. If we were to expand to four elements and consider quaternary compounds, this number expands to a magnitude of over 32 billion possible combinations[11]. Hence, it is necessary to have a down-selection algorithm that can pick up high-potential candidate to prioritize the research effort towards the most promising compounds. To achieve this, we utilize the concept of oxidation states as a means of accelerating material discovery.

In literature, there are claims that machine-learned models for the formation energy of compounds can be as accurate as Density Functional Theory (DFT). High throughput DFT databases like the Open Quantum Materials Database (OQMD) contain over 200,000 DFT calculated crystal structures[12]. OQMD can be used in training a machine learning model to predict new stable ternary compounds. Bartel *et al*. tested seven machine learning models[13]. They found that while formation energies can be predicted accurately, all the compositional models tested performed poorly in predicting the stability of compounds, making them less effective compared to DFT for discovering and designing new solids. Nandy *et al*. used natural language processing-based procedure to mine literature related to metal-organic frameworks (MOFs), and presented a workflow to predict the stability of structurally characterized MOFs based on solvent removal and thermal stabilities[14]. They tested the model on unseen data and correctly predicted the stability of 31 out of 40 MOFs. In 2016, Davies *et al*. introduced a framework (SMACT) that filters chemically implausible compositions by applying principles of charge neutrality and electronegativity[11]. Based on this, the same team demonstrated two years later that by combining machine learning, and statistical analysis of oxidation states, it is possible to construct a high-throughput search for new stable ternary materials[15]. Specifically, the study focused on halide compounds that contain at least one metal element and one of the anions of interest and used the model to explore these compounds.

Unfortunately, only a small fraction of the candidates proposed by these frameworks have been validated through experimental synthesis[16]. This is because materials synthesis requires not just the correct composition, but also the correct synthesis pathway, the discernment of which is not trivial. The existing frameworks lack the ability to provide these methods, hence the difficulty in synthesizing many suggested compounds. As a result, there is a need to identify the most promising candidates that can be synthesized with current synthesis capabilities. Adopting an approach that relates the suggested compounds to those that have already been synthesized is crucial for finding the next compound which is most likely to be synthesized.

Material properties are manifested in a large part through atomistic structures and crystal symmetries, which can be analysed through software packages such as *crystal-toolkit*[17]. However, implementing crystal structure analytics requires specifying the spatial coordinates of each atom, making it unfeasible in our case where a large number of hypothetical compositions are generated without reference to atomic coordinates. As such, it is a more feasible first step to eliminate unphysical compositions based on composition alone.

In this work, we formulate an approach of ranking the synthesizability of ternary compositions using oxidation state probability (OSP), a metric for evaluating a hypothetical ternary composition. OSP is built upon the principle that the probability that each element in a ternary compound can be assigned a specific oxidation state, which can be extracted from the Materials Project (MP) database. Based on this, an arbitrarily large pool of hypothetical ternary compositions can be evaluated at scale. Unlike SMACT[15], OSP does not calculate probabilities based on pairwise atomic bonding, but simply uses the product of elemental probabilities. In the context of a specific material system, this simplifies the computational process. Furthermore, our method can be generalized to material systems with more than 3 elemental components.

Other state-of-the art ranking methods, including formation energy above hull and machine learning-trained synthesizability score, are compared to the OSP metric. We select a ternary system, Cu-In-Te, which has shown promising thermoelectric properties, to test the validity of our approach. We generate the potential compounds in the ternary system and rank them according to the OSP filter. Finally, we evaluate the performance of the OSP filter by comparing its results with existing literature.

# Methodology

Generation of elemental oxidation state probability

Before creating the OSP model, we first establish a database which assigns oxidation states to elements for various compounds. The process begins by selecting a ternary system of interest to serve as the subject of study. Next, the Materials Project (MP) database is queried to identify compounds that are composed of these elements, and the results are then screened using the Inorganic Crystal Structure Database (ICSD). MP database entries attributed to a corresponding ICSD entry are deemed to be *attested* (naturally-occuring or synthesized, as opposed to purely theoretical), and are selected for further analyses. Note that our criterion for attested materials is restricted to the corroboration between the MP and ICSD databases, and exclude other materials which may have been discovered, but are not yet included in these databases at the time of writing.

To determine the possible oxidation states for each compound, we compile a charge pool that encompasses all known oxidation states for each element, along with their prevalence. Thus, for each given element X bearing an oxidation state of $c$ ($c$ = -5, -4, -3, …, +8, or +9), we obtain a measure of prevalence

$$P(X^c) = \frac{\#(\text{compounds where } X \text{ assumes oxidation state } c)}{\#(\text{all attested compounds containing } X)}$$

representing the global probability of the element X in adopting the oxidation state $c$ in a compound. A complete table of $P(X^c)$ concerning elements included in our analysis is presented in the supplementary materials (Table S1).

As a constraint, the net oxidation state of a compound ($Q$), determined by the sum of the individual oxidation states of each element present in the compound multiplied by the corresponding number of atoms, is required to be zero before the compound is to be accepted as valid. In the example of a ternary compound $(A^q)_x(B^{q'})_y(C^{q''})_z$, where $q$, $q'$, and $q''$ denote oxidation states of A, B, and C respectively, we choose a possible combination of oxidation states so that the net charge

$$Q(A_x B_y C_z) = xq + yq' + zq'' \equiv 0,$$

thus fulfilling the condition of charge neutrality. In addition, each ternary oxidation state configuration is assigned a OSP metric $\mathrm{OSP}(A^q B^{q'} C^{q''})$, a measure of joint probability defined as a product of the probabilities $P(A^q), P(B^{q'})$, and $P(C^{q''})$ of the individual elements

$$\mathrm{OSP}(A^q B^{q'} C^{q''}) = P(A^q) \times P(B^{q'}) \times P(C^{q''}).$$

In the case where more than one combination of $q$, $q'$, and $q''$ satisfies charge neutrality for a composition $(A^q)_x(B^{q'})_y(C^{q''})_z$, the composition is excluded in subsequent steps. This is done to simplify analysis.

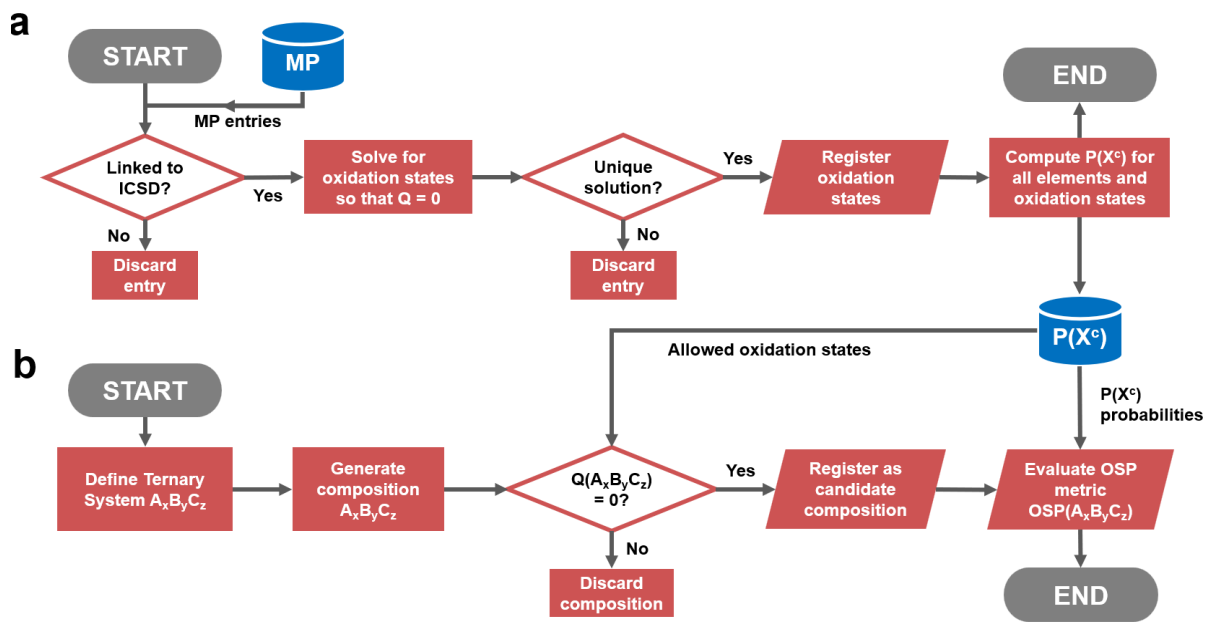

*Figure 1. The processes of (a) solving and compiling oxidation state prevalence figures for elements and (b) generating and evaluating candidate composistions for a given ternary system.*

Oxidation states are assigned with reference to Pauling electronegativity values for the elements: the element with the highest electronegativity is assigned negative oxidation states exclusively, while the element with the least electronegativity is given positive oxidation states only[18]. The middle element may be assigned both negative and positive oxidation states. Within user-specified total atom limitations, charge neutral compounds are produced by ensuring the total charge is zero. The flow chart in Figure 1 summarizes the process of generation of material compositions satisfying the charge neutrality condition.

# Results and Discussion

## Benchmarking

To establish the efficacy of our code in determining the relative likelihood of different compounds, we subject the OSP filter to a benchmarking process using 1000 ternary systems. The ternary systems are ranked according to the number of MP entries attributed to the systems, in descending order. The top-ranked ternary system (Fe-O-F) contains 520 MP entries, while the 1000[th] system (Co-Sn-Tb) contains 7 entries. We assume that the ternary systems in question have been thoroughly searched and the most probable compound for each ternary system has already been identified. As such, we require that the ternary systems chosen for benchmarking possess a large amount of MP entries. Next, for a given ternary system, each MP compound is assigned a OSP metric, and ranked in descending order of the OSP metric. The highest-ranked attested MP compound is noted and its rank is registered as the *rank advantage* of the ternary system. A lower integer value (ideally 1) of rank advantage indicates that the OSP metric is successful in predicting synthesizable compositions in the ternary system. The evaluated rank advantages of 1000 ternary systems, as well as the OSP of the highest-ranked compound with a corresponding ICSD entry, are shown in Figure 2a.

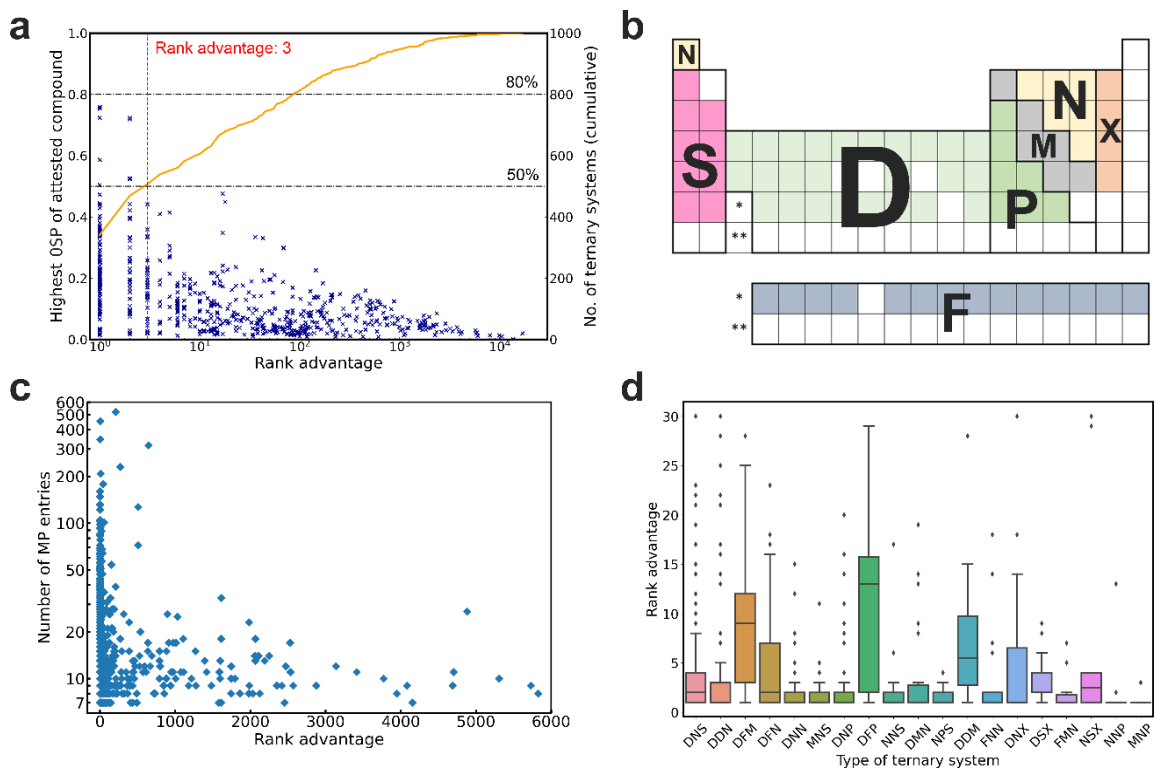

Figure 2. a) The rank advantages of the 1000 ternary systems referred in the benchmarking process. The scatter plot points are the oxidation state probability (OSP) of the highest-ranked attested compound each ternary system, while the yellow line represents the cumulative count of ternary systems against the rank advantage. b) Classification scheme for elements used to describe the chemical type of ternary systems, Elements which are featured in our benchmarking analysis are shaded. c) Relationship between rank advantage and the number of MP entries of the ternary systems. d) Distribution of rank advantage in benchmarking, grouped by the chemical type of ternary system.

Approximately half of the 1000 ternary systems are assigned the rank advantage of 3 or below by the OSP metric. However, about 20% of the ternary systems exhibit rank advantages beyond 100, revealing that the metric may not perform as well for these systems. There are two possible explanations behind this phenomenon. Firstly, if the nature of interatomic bonding is closer to metallic bonding than to ionic or covalent bonding, the notion of oxidation state is rendered irrelevant, leading to inaccurate assignment of oxidation states. Secondly, it is possible that these ternary systems simply have not been well-explored at the time of writing.

We now aim to test the hypothesis that the OSP metric performance depends on the chemical nature of the elemental pairings in the ternary system. To this end, we organize the ternary systems according to the classification of its component elements[19], where each element is assigned a symbol according to its chemical type: **n**onmetal (N), **s**-block (alkali and alkali earth) metal (S), **m**etalloid (M), halogen (X), **p**-block metal (P), **d**-block (transition) metal (D), and **f**-block or lanthanide (F) (Figure 2b). A combination of 3 symbols denotes the chemical type of the ternary system. For example, the symbol DFP would represent a ternary system containing one transition metal, one lanthanide, and one p-block metal, whereas NNX would denote a combination of two nonmetal elements and a halogen.

Figure 2c depicts the distribution of rank advantages for all ternary system types. The greatest variability in rank advantages occur when the ternary system has D, F, or X elements. Conversely, the rank advantages are consistently higher if the ternary system type contains N, S, or P elements, indicating better performance of the OSP metric. (The same analysis, broken down by element, is presented in the supplementary information in Figure S2.) Additionally, the higher index numbers may be due to the fact that some ternary systems haven't been thoroughly explored rather than a decrease in the performance of the filter[20]. This is supported by the observation that ternary systems with lower rank advantage typically have lower numbers of MP entries attributed to them (Figure 2d). For example, all ternary systems with a rank advantage of >1000 are represented by at most 40 MP entries.

The poorer performance of the OSP method in halogen-containing systems can be surprising, since the higher electronegativities of halogens would imply covalent or ionic bonding over metallic bonding in the material. This could be attributed to the higher variability of oxidation states possible for nonmetal p-block and d-block transition metals.

Comparison with alternative methods

We now apply the OSP method to the Cu-In-Te ternary system, which includes the well-known thermoelectric material $CuInTe_2$[21] and can potentially harbor high-performing thermoelectric compounds. Among the hypothetical Cu-In-Te compositions, the 4 highest-ranked compositions by OSP metric are all assigned the OSP metric value of 7.01%, corresponding to $Cu^+$ (41.4% prevalence), $In^{3+}$ (36.8% prevalence), and $Te^{2-}$ (46.0% prevalence). These are: $CuInTe_2$, $CuIn_3Te_5$, $Cu_3InTe_3$, and $Cu_5InTe_4$.

Two existing ranking systems were included in this study to compare with the OSP method. The first is based on formation energy above the convex hull by using the *Roost* (**R**epresentati**o**n Learning fr**o**m **St**oichiometry) machine learning model that uses deep representation learning to predict the synthesizability score of a potential candidate at a low computational cost [4]. The algorithm represents the stoichiometric formula as a dense weighted graph between elements. The model was trained using ICSD compounds, including single element, binary, and ternary compounds, and achieved a mean average error of approximately 20 meV/atom. The predicted formation energy is then used to calculate the energy above the hull by comparing it with the existing hull surface in the MP database, and materials with a lower formation energy relative to the hull is deemed more synthesizable.

The second filter is based on a machine learning algorithm that predicts the synthesizability score of a potential candidate compound based on its Fourier-transformed crystal properties (FTCP) representation[22]. This algorithm provides a filter for new materials and offers an efficient and accurate way to classify and identify promising material candidates. The algorithm is also trained on ICSD compounds and serves as the filter to assess the synthesizability of the compounds. This SC algorithm provides a filter for new materials and offers an efficient and accurate way to classify and identify promising material candidates.

*Table 1. The top 4 most likely compounds that can be synthesized are ranked by oxidation state probability (OSP) metric. The score of the other two models: Fourier-transformed crystal properties (FTCP)-based synthesizability scores [22] and formation energy above the hull ($E_{hull}$) from the Roost model[4] are shown.*

| Rank | Composition | OSP (%) | Roost $E_{hull}$ (meV/at) | FTCP score (%) |
|------|-------------|---------|---------------------------|----------------|
| 1    | $CuInTe_2$  | 7.01%   | -4.8                      | 98.9%          |
| 2    | $CuIn_3Te_5$| 7.01%   | -157.5                    | 9.5%           |
| 3    | $Cu_3InTe_3$| 7.01%   | 63.5                      | 96.3%          |
| 4    | $Cu_5InTe_4$| 7.01%   | 95.0                      | 55.8%          |

Table 1 lists the most synthesizable compositions of Cu-In-Te based on the OSP metric, as well as the evaluation of the same compositions using *Roost* and FTCP. We note the variability of *Roost* and FTCP synthesizability scores in contrast with the simplicity of the OSP metric, which only considers oxidation states. For the attested material $CuInTe_2$, *Roost* indicates a negative $\Delta E_F$ (-4.8 meV/at), and FTCP indicates a relatively high score of 98.9%, indicating good agreement among the three methods. In the other compositions, the agreement is less apparent.

Synthesis and characterization of proposed compounds

The 4 proposed $Cu_xIn_yTe_z$ compounds were synthesized, with $CuInTe_2$ being used as a reference compound. $Cu_xIn_yTe_z$ samples were synthesized using the solid-state reaction method. High purity (Macklin 99.99% metal basis) electrolytic copper powder, indium ingots, and tellurium ingots were used. Stoichiometric amounts of each element were weighed and transferred into a cleaned quartz ampoule. The quartz ampoules holding the samples were then vacuum sealed at a pressure of ≈$10^{-3}$ Pa and annealed in a furnace at 900°C for 36h. The ramp up and cooling rates were set at 1.8°C min$^{-1}$. Annealed ingots are subsequently grinded in an agate mortar into a fine powder and loaded into a graphite die. They are then sintered using the spark plasma sintering machine (ELENIX, EdPas) at 580°C, 13MPa compression for 15 mins. The resulting product is a highly compact cylindrical pellet. Part of the pellet is then grinded down into a fine powder again using the agate mortar for subsequent characterization.

Scanning electron microscopy (SEM) was conducted using a JEOL 7800F field-emission SEM equipped with an Oxford INCA EDX detector. 3 maps at different length scales and at different sites were acquired for at least three minutes to probe elemental homogeneity across the films and powders. Cu, In, and Te distributions in the sample appear to be homogeneous (Figure S3). The elemental composition of the sample, averaged across the 3 maps, is found to be 11.8 at. % Cu, 32.5 at. % In, and 51.1 at. % Te, close to the projected ratio of 1:3:5. This result was used to facilitate the Rietveld refinement following the X-ray Diffraction (XRD) measurement.

XRD analysis was carried out on a Bruker D8 Advanced diffractometer under coupled θ-2θ geometry with a Cu Kα X-ray source operated at 40kV and 40mA ($\lambda_{CuK\alpha}$ = 1.54056 Å) and a beam knife for low-background low-angle signal. $CuIn_3Te_5$ is confirmed after XRD analysis and Rietveld refinement to be a pure phase compound (Figure 3, Table 2). The figures of merit are: $R_{exp}$ = 5.71, $R_{wp}$ = 13.21, $R_p$ =

9.62, and GOF = 2.31. Rietveld refinement reveals that the stable phase of CuIn$_3$Te$_5$ to belong to the space group $P\bar{4}2c$ (space group number 112). The phase is tetragonal, with the lattice parameters $a$ = 6.164Å and $c$ = 12.340Å. Our characterization of the crystal structure agrees well with earlier studies on CuIn$_3$Te$_5$, such as Díaz et al.[23] and later thin film studies by Mise and Nakada [24]. The latter noted the general disposition of the CuIn$_3$Te$_5$ crystal structure as a defect (ordered vacancy) chalcopyrite. Despite the concerns of Marín et al.[25] that ambiguities arising from the interpretation of weak reflections in the XRD analysis can give rise to multiple possible interpretations on crystal symmetry, the variations of interpreted space groups simply relate to the differences in the occupancies of specific Wyckoff sites, rather than differences in the underlying structural motif which is defect chalcopyrite. Furthermore, the closely related CuIn$_3$Se$_5$ was found to be of the same structure[26].

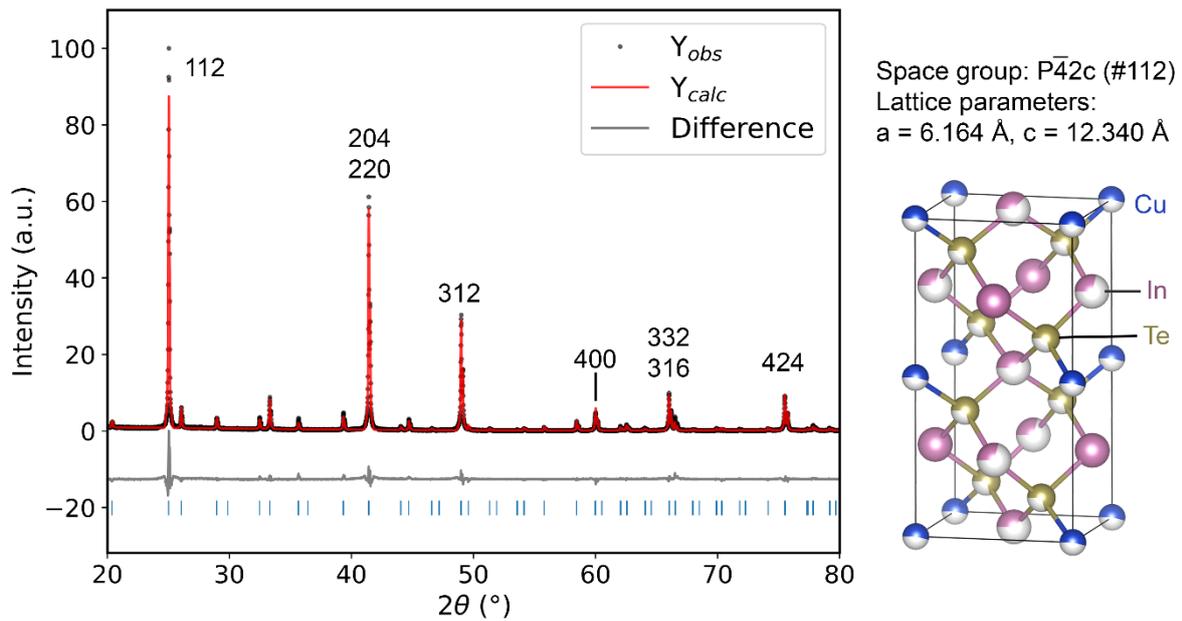

*Figure 3. XRD peak profile of single-phase CuIn$_3$Te$_5$ obtained through experimental synthesis and Rietveld refinement. The characteristic Bragg peaks are labelled.*

*Table 2. Fractional atomic site coordinates (x, y, z) and occupancies (Occ.) in the P-chalcopyrite CuIn$_3$Te$_5$ unit cell derived from Rietveld refinement*

| Crystal system: Tetragonal, Space Group: $P\bar{4}2c$ (No. 112) | | | | | |
|---|---|---|---|---|---|
| $a = b = 6.164$ Å, $c = 12.340$ Å, $V = 468.896$ Å$^3$ | | | | | |
| Atom | Wyckoff site | Occ. (%) | x | y | z |
| Cu | 2e | 55.8 | 0 | 0 | 0 |
| In$_1$ | 2b | 99.6 | 0.5 | 0 | 0.25 |
| In$_2$ | 2f | 38.7 | 0.5 | 0.5 | 0 |
| In$_3$ | 2d | 28.4 | 0 | 0.5 | 0.25 |
| Te | 8n | 69.5 | 0.2331 | 0.2580 | 0.1173 |

The structure of CuIn$_3$Te$_5$ determined by our analysis is not hitherto represented in either MP or ICSD databases. From this discovery, we receive confirmation that the OSP method was successful in predicting the synthesis of a material outside of the training set. On the other hand, the attempt syntheses of Cu$_3$InTe$_3$ and Cu$_5$InTe$_4$ were less successful, generating mixtures of CuInTe$_2$ and Cu$_{1.75}$Te instead of a single-phase product (Figure S4). After Rietveld refinement, we concluded that the synthesis products for these compositions are 86.4% CuInTe$_2$ / 13.6% Cu$_{1.75}$Te for Cu$_3$InTe$_3$, and 75.4% CuInTe$_2$ / 24.6% Cu$_{1.75}$Te for Cu$_5$InTe$_4$.

To further illustrate the performance of the filter beyond the training data, we analyzed the first 30 ternary systems which were ranked by their number of MP entries in descending order. All these 30 ternary systems are found to have a rank advantage of 1. We then manually searched the literature for the three top-ranked compounds deemed non-attested based on the lack of a corresponding ICSD-linked MP entry. Several materials turn out to be reported in literature (Table 3) even though they were excluded from the training dataset for this algorithm. This strongly suggests that the OSP method is also capable of predicting existing compounds when such compounds are excluded from prior knowledge.

*Table 3. These seven compounds are found in the top 3 ranked of the list without ICSD ID compounds in their respective ternary systems ranked by the oxidation state probability filter. 30 ternary systems with the greatest number of MP entry were explored.*

| Compound | Rank | Reference |
| --- | --- | --- |
| $Li_3V_2O_5$ | 3 | [27] |
| $Mg_2Fe_2O_5$ | 1 | [28] |
| $Na_2Ti_2O_5$ | 1 | [29] |
| $Ca_3Fe_2O_5$ | 3 | [30] |
| $Mg_2TiO_4$ | 1 | [31] |
| $LiNb_2O_6$ | 1 | [32] |
| $Fe(OH)_3$ | 1 | [33] |

# Conclusion

Our study presents an algorithm that assigns a measure of prevalence of a certain oxidation state to a certain element in the form of a probability. After compiling the probabilities from a training set drawn from the MP and ICSD databases, we were able to establish a OSP metric which enables us to sift through a large number of procedurally-generated compositions for those which are the most plausible, i.e. those who bear the greatest likelihood of being synthesizable or being found in nature. Through a comparison of the three main ranking methods, namely *Roost*, OSP, and FTCP synthesizability score, our analysis found that the OSP method showed the most promising results for the Cu-In-Te ternary system. We were able to synthesize and characterize a single-phase sample of $CuIn_3Te_5$, a candidate compound proposed by the OSP method. In addition, the OSP method managed to decuce the occurrence of several compounds not included in the original training set. Our method has the potential to accelerate the discovery of novel compounds by providing a guide for experimentalists to easily select from a pool of thousands of candidates. We demonstrated the potential of the OSP metric for filtering and identifying several promising candidate compounds for further experimental investigation. This approach to materials design can be combined with machine learning algorithms and generative design frameworks to accelerate the discovery of new materials.